\documentclass[%
 reprint,
 showkeys,
nofootinbib,
 amsmath,amssymb,
 aps,
]{revtex4-2}

\usepackage{graphicx}
\usepackage{dcolumn}
\usepackage{bm}
\usepackage{algpseudocode}
\usepackage{hyperref}
\usepackage{wasysym}

\hypersetup{
     colorlinks = true,
     linkcolor = red,
     citecolor = green,
     urlcolor = blue
}

\begin{document}

\preprint{APS/123-QED}

\title{Robust Bayesian non-linear pulsar timing and noise analysis using a sequential data-tempered ensemble sampler}

\author{Abhimanyu Susobhanan}
 \email{abhimanyu@iisertvm.ac.in}
\affiliation{School of Physics, Indian Institute of Science Education and Research Thiruvananthapuram, Maruthamala P.O., Thiruvananthapuram, Kerala 695551, India}%

\date{\today}

\begin{abstract}
We develop a sequential ensemble MCMC scheme for Bayesian pulsar timing and noise analysis that uses data size as the tempering parameter. The method enables gradual and reliable convergence from potentially uninformed initial points by iteratively increasing the data size, thereby increasing sensitivity and progressively narrowing the target distribution until the posterior of the full dataset is reached. The initial iterations use sub-datasets that can be orders of magnitude smaller than the full dataset, substantially reducing computational cost, which is further lowered by employing approximate lower-rank models justified by the reduced sensitivity of these sub-datasets. Bayesian inference is performed using the final MCMC chain sampled from the true target distribution after applying appropriate burn-in. We demonstrate the method using the NANOGrav 12.5-year data of PSR B1855+09, showing progressive convergence to the target posterior and that most of the computational time is spent sampling the true target distribution rather than in the preceding tempered iterations.
\end{abstract}

\maketitle


\section{Introduction}
\label{sec:intro}

Pulsars are rotating neutron stars whose electromagnetic radiation appears as periodic pulses to a terrestrial observer \citep{LorimerKramer2012}.
Their high rotational stability enables us to use them as celestial clocks to probe a wide variety of time-domain phenomena, from gravitational waves (GWs) \citep{AgazieAntoniadis+2024} and neutron star equation of state \citep[e.g.][]{CromartieFonseca+2020} to solar wind \citep[e.g.][]{TiburziShaifullah+2021}.
The technique underlying these applications is pulsar timing, which allows us to track the rotation of the pulsars by measuring the times of arrival (TOAs) of their pulses \citep{HobbsEdwardsManchester2006}.
Pulsar Timing Arrays (PTAs) are experiments that detect GWs in the nanohertz frequency regime by monitoring ensembles of millisecond pulsars distributed across the sky, thereby synthesizing a Galaxy-sized GW detector via pulsar timing \citep{FosterBacker1990}.
Recently, PTA experiments found evidence for a stochastic GW background, laying the foundations of nanohertz GW astronomy \citep{AgazieAntoniadis+2024}.

The measured TOAs of a pulsar are influenced by several astrophysical and instrumental processes, some of which are deterministic while the others are stochastic. 
A mathematical description of these deterministic processes is known as the pulsar timing model, and a description of the stochastic processes is known as the pulsar noise model.
The timing model parameters are usually estimated in a frequentist setting using packages such as \texttt{tempo2} \citep{HobbsEdwardsManchester2006,EdwardsHobbsManchester2006} and \texttt{PINT} \citep{LuoRansom+2021,SusobhananKaplan+2024} assuming fixed noise parameters, whereas the noise model parameters are estimated in a Bayesian setting using packages such as \texttt{ENTERPRISE} \citep{JohnsonMeyers+2024} and \texttt{DISCOVERY} \citep{VallisneriMeyers+2025} using a linear approximation of the timing model.
Bayesian inference using the full non-linear timing and noise model can be done using \texttt{TEMPONEST} \citep{LentatiHobson+2014} and \texttt{Vela.jl} \citep{Susobhanan2025a,Susobhanan2025b}.
The linear approximation of the timing model enables the timing model parameters to be analytically marginalized \citep{van_HaasterenLevin2013} or Gibbs sampled \citep{LaalLamb+2023}.
Recently, \citet{GundersenVan_Haasteren+2026} introduced a standardizing parameter transformation to improve the efficiency of Bayesian inference for the linearized timing and noise model.
These methods significantly reduce the computational cost of Bayesian inference, and pulsar noise analysis is usually done assuming the linearized timing model for this reason.

However, the pulsar timing model is intrinsically non-linear and the linearized timing model does not always approximate it accurately.
Further, some timing model parameters require prior distributions which are not conducive to analytic marginalization or Gibbs sampling.
This is especially true for pulsar timing applications where the astrophysics of interest is encoded in the non-linear timing parameters, such as tests of theories of gravity \citep[e.g.][]{VaglioCarleo+2026}.
In such circumstances, it is necessary to perform Bayesian inference over the full non-linear timing and noise model.
We also note that in most cases it is not necessary to treat \textit{all} parameters non-linearly; some parameters, such as the rotational frequency and its derivatives, can indeed be treated as linear parameters without loss of accuracy \citep{Susobhanan2025b}.

There are three major computational challenges in performing Bayesian inference using the non-linear timing and noise model.
First, the non-linear timing model involves expensive mathematical operations, including numerical solution of coupled differential equations in some cases \citep[e.g.][]{ArchibaldGusinskaia+2018}.
Secondly, the timing and noise model can have a large number of parameters, up to hundreds in some cases, depending on the complexity of the model and the dataset.
Finally, the pulsar timing posterior distribution often features complex  geometries which can impede the convergence of sampling algorithms.

In this paper, we develop a new sequential data-tempered ensemble sampling algorithm to efficiently perform non-linear pulsar timing and noise analysis, addressing the above concerns.
This paper is arranged as follows.
In Section \ref{sec:model}, we briefly describe the pulsar timing and noise model, the pulsar timing posterior distribution, and its computational cost.
We introduce the new sampling scheme in Section \ref{sec:sdt}, and demonstrate it using PSR B1855+09 in Section \ref{sec:demo}.
Finally, we summarize our results in \ref{sec:summary}.
Please note that we restrict our discussions to the narrowband timing paradigm in this work, but the sampling method presented here is equally applicable to the wideband timing paradigm \citep{PennucciDemorestRansom2014}. 

\section{The pulsar timing and noise model}
\label{sec:model}
We begin by summarizing the pulsar timing and noise model, and the corresponding posterior distribution.
In pulsar timing, the fundamental measurable quantity is the TOA $t_i$ of a pulse recorded at an observatory \citep{Taylor1992}.
$t_i$ is related to the corresponding  pulse emission time $\tau_i$ by the expression
\begin{equation}
    t_i = \tau_i + \Delta(t_i) + \mathcal{N}_R\,,
\end{equation}
where $\Delta(t_i)$ is the sum of all astrophysical and instrumental delays influencing the measured TOA, and $\mathcal{N}_R$ is a stochastic process arising from the radiometer noise of the telescope.
The rotational phase $\phi_i$ of the pulsar can be estimated from $\tau_i$ using the expression
\begin{equation}
    \phi_i = \phi_0 +\sum_{j=0}^{N_F} \frac{F_j(t_i-t_0)^{j+1}}{(j+1)!} + \Phi(t_i) + \phi_\text{ARN}(t_i) + \mathcal{N}_J\,,
\end{equation}
where $\phi_0$ is an initial phase, $F_n$ represent the rotational frequency of the pulsar and its derivatives, $\Phi(t_i)$ represents the various phase corrections, $\phi_\text{ARN}(t_i)$ represents the slow stochastic wandering of the apparent rotational frequency known as the achromatic red noise, and $\mathcal{N}_J$ is a stochastic term representing pulse jitter. 
Detailed descriptions of the various physical processes contributing to the delays and phase corrections mentioned above may be found in \citet{HobbsEdwardsManchester2006,EdwardsHobbsManchester2006,LentatiHobson+2014,LuoRansom+2021,Susobhanan2025a}, and references therein.

The timing residual $r_i$ is given by
\begin{equation}
    r_i = \frac{\phi_i - N[\phi_i]}{\bar{F}_i}\,,
\end{equation}
where $N[\phi_i]$ is the integer closest to $\phi_i$ and $\bar{F}_i$ is the topocentric frequency of the pulsar.
The pulsar timing log-likelihood function can be written in terms of $r_i$ as \citep{LentatiHobson+2014,Susobhanan2025a}
\begin{equation}
    \ln L[\mathfrak{D}|\boldsymbol{\theta},\mathfrak{M}] = -\frac{1}{2}\textbf{r}^T \textbf{C}^{-1} \textbf{r} - \frac{1}{2}\ln\det[2\pi \textbf{C}]\,,
    \label{eq:lnlike}
\end{equation}
where $\mathfrak{D}$ is the dataset containing the TOA measurements, their uncertainties, and related metadata, $\mathfrak{M}$ is the pulsar timing and noise model, $\textbf{r}$ is a vector containing the residuals $r_i$, $\boldsymbol{\theta}$ is the collection of model parameters, and $\textbf{C}$ is the TOA covariance matrix representing the various noise processes present in the data.
$\textbf{C}$ is usually expressed in a reduced-rank form \citep{van_HaasterenVallisneri2014a}
\begin{equation}
    \textbf{C} = \textbf{N} + \textbf{F} \boldsymbol{\Phi} \textbf{F}^T\,,
\end{equation}
where $\textbf{N}$ is a block-diagonal matrix quantifying the radiometer noise and the pulse jitter, 
and $\textbf{F}$ and $\boldsymbol{\Phi}$ are respectively the reduced-rank basis matrix and the prior covariance matrix representing the time-correlated stochastic processes such as spin noise and interstellar dispersion measure variations. 
Finally, the posterior distribution can be written using the Bayes theorem as
\begin{equation}
    \ln P[\boldsymbol{\theta}|\mathfrak{D},\mathfrak{M}] =  \frac{L[\mathfrak{D}|\boldsymbol{\theta},\mathfrak{M}] \; \Pi[\boldsymbol{\theta}|\mathfrak{M}]}{ \mathcal{Z}[\mathfrak{D}|\mathfrak{M}]}\,,
\end{equation}
where $ \Pi[\boldsymbol{\theta}|\mathfrak{M}]$ represents the prior distributions of the model parameters, and $\mathcal{Z}[\mathfrak{D}|\mathfrak{M}]$ is the Bayesian evidence.
Bayesian parameter estimation of the pulsar timing and noise model involves drawing samples from $P[\boldsymbol{\theta}|\mathfrak{D},\mathfrak{M}]$, using Markov Chain Monte Carlo (MCMC) or nested sampling methods.

\subsubsection{The computational cost of the pulsar timing and noise model}
Consider a dataset with $n$ TOA measurements with a $p$-rank covariance matrix such that $p\ll n$. 
In this case, $\textbf{r}$ is an $n$-dimensional vector, $\textbf{N}$ is an $n\times n$ block-diagonal matrix, $F$ is an $n\times p$ matrix, and $\boldsymbol{\Phi}$ is a $p\times p$ matrix.
Using the Woodbury identity, $\textbf{r}^T\textbf{C}^{-1} \textbf{r}$ can be computed as follows: 
\begin{equation}
  \textbf{r}^T\textbf{C}^{-1} \textbf{r} = \textbf{r}^T\textbf{N}^{-1} \textbf{r} - \textbf{b}^T \textbf{b}\,,
\end{equation}
where
\begin{subequations}
\begin{align}
    \textbf{b} &= \textbf{L}^{-1} \textbf{y}\,,\\
    \textbf{y} &= \textbf{F}^T \textbf{N}^{-1} \textbf{r}\,,\\
    \textbf{L} \textbf{L}^T &= \boldsymbol{\Sigma}^{-1}= \boldsymbol{\Phi}^{-1} + \textbf{F}^T \textbf{N}^{-1}\textbf{F}\,,
\end{align}
\end{subequations}
where $\textbf{L}$ is the lower-triangular Cholesky factor of $\boldsymbol{\Sigma}^{-1}$\,.
Similarly, using the matrix determinant lemma, we have
\begin{equation}
    \ln\det\textbf{C} = \ln \det \textbf{N} + \ln \det  \boldsymbol{\Phi} + 2 \ln\det \textbf{L}\,.
\end{equation}

Assuming a sufficiently large $n$ and an appropriate block-diagonal structure for $\textbf{N}$ \citep{JohnsonMeyers+2024},
this computation is dominated by $\textbf{F}^T \textbf{N}^{-1} \textbf{F}$, which has a time complexity of $\mathcal{O}(np^2)$.
We also note that larger datasets require higher-rank representations of noise in practice, making $p$ an increasing function of $n$. 
Therefore, the computational complexity of $\ln P[\boldsymbol{\theta}|\mathfrak{D},\mathfrak{M}]$ scales super-linearly with $n$ for real datasets.

The computational challenge in performing Bayesian inference over $P[\boldsymbol{\theta}|\mathfrak{D},\mathfrak{M}]$ arises from three factors, as mentioned in the introduction.
The first is of course the $\mathcal{O}(np^2)$ computational cost mentioned above, which grows larger for larger datasets\footnote{This is assuming sufficiently large values of $n$ and $p$. In practice, the computational cost may be dominated by the $\mathcal{O}(n)$ evaluation of $\textbf{r}$ if the timing model is complex, especially when parts of the timing model are not tractable analytically.}.
Secondly, large pulsar timing datasets often include TOAs obtained using multiple observing systems with varying noise properties, requiring many noise and instrumental delay parameters.
Further, larger datasets are sensitive to detect more number of astrophysical parameters, which, together with the noise parameters, lead to large-dimensional parameter spaces. 
Finally, the posterior distribution usually features a sharply peaked global maximum occupying only a tiny portion within the prior volume, often accompanied by Neal's funnel-like geometry \citep{Neal2003,GundersenVan_Haasteren+2026}.
These factors often lead to slow convergence while using MCMC and nested sampling, both in terms of time as well as the number of posterior evaluations.

\section{The new sampling method}
\label{sec:sdt}

Consider a proper subset of the data $\mathfrak{D}' \subset \mathfrak{D}$. 
We expect the distribution $P[\boldsymbol{\theta}|\mathfrak{D}',\mathfrak{M}]$ to be (a) cheaper to compute than $P[\boldsymbol{\theta}|\mathfrak{D},\mathfrak{M}]$, and (b) broader than $P[\boldsymbol{\theta}|\mathfrak{D},\mathfrak{M}]$ because less data leads to a less constrained posterior distribution.
Further, since the dataset $\mathfrak{D}'$ is less sensitive than $\mathfrak{D}$, we can approximate $\mathfrak{M}$ using a lower-ranked model $\mathfrak{M'}$, i.e., $P[\boldsymbol{\theta}|\mathfrak{D}',\mathfrak{M}']\approx P[\boldsymbol{\theta}|\mathfrak{D}',\mathfrak{M}]$.
The computational cost of evaluating $P[\boldsymbol{\theta}|\mathfrak{D}',\mathfrak{M}']$ is asymptotically $\mathcal{O}(n'{p'}^2) < \mathcal{O}(np^2)$, where $n'$ is the number of TOAs in $\mathfrak{D}'$ and $p'$ is the rank of the correlated noise representation in the approximate model $\mathfrak{M}'$.

We now define a sequence of progressively larger non-empty datasets 
$$\mathcal{D}_1 \subset \mathcal{D}_2 \subset  ... \subset \mathcal{D}_S\equiv\mathcal{D},$$
and their corresponding approximate lower-rank models 
$$\mathcal{M}_1 \subset \mathcal{M}_2 \subset... \subset \mathcal{M}_S\equiv\mathcal{M}.$$
We choose the data-subsets such that the number of TOAs $n_s\sim \xi^s n$, where $\xi\in(0,1)$, and the approximate models $\mathcal{M}_i$ such that $p_i \sim n_i p/n$, while ensuring that each data-subset $\mathcal{D}_s$ contains a minimum number of TOAs for each observing system.
Then, the computational cost of $P[\boldsymbol{\theta}|\mathfrak{D}_s,\mathfrak{M}_s]$ is cheaper than $P[\boldsymbol{\theta}|\mathfrak{D}_{s+1},\mathfrak{M}_{s+1}]$ by a factor of $\xi^3$ in the limiting case when $n$ is large.
i.e., $P[\boldsymbol{\theta}|\mathfrak{D}_s,\mathfrak{M}_s]$, where $s<S$, is significantly easier to sample from than $P[\boldsymbol{\theta}|\mathfrak{D},\mathfrak{M}]$ due to the reduced cost of its evaluation and its broader shape.

Based on the above observations, we propose a sequential data-tempered ensemble MCMC scheme as follows.
\begin{enumerate}
    \item Given a set of $W$ number of starting points $\boldsymbol{\theta_{w0}}$, the posterior distributions $P[\boldsymbol{\theta}|\mathfrak{D}_s,\mathfrak{M}_s]$, burn-in factors $\beta_s$, and number of steps $l_s$ for each $s$,
    \item for $s \in \{1,2,...,S\}$
    \begin{enumerate}
        \item Run $l_s$ number of ensemble MCMC steps on $P[\boldsymbol{\theta}|\mathfrak{D}_s,\mathfrak{M}_s]$ starting with $\boldsymbol{\theta_{w0}}$.
        \item Replace $\boldsymbol{\theta_{w0}}$ by randomly drawing $w$ \textit{unique} samples from the resulting ensemble MCMC chain after discarding the first $\beta_s l_s$ steps.
    \end{enumerate}
    \item Return the final ($s=S$) MCMC chain after discarding $\beta_S l_S$ steps.
\end{enumerate}
Here, we implement the ensemble MCMC steps (2a) using the \texttt{emcee} package \citep{Foreman-MackeyHogg2013}, with a combination of the stretch move \citep{GoodmanWeare2010} and the differential evolution move \citep{NelsonFordPayne2014}. 

Note that the computational cost of evaluating $P[\boldsymbol{\theta}|\mathfrak{D}_1,\mathfrak{M}_1]$ is negligible compared to $P[\boldsymbol{\theta}|\mathfrak{D},\mathfrak{M}]$, provided $\xi$ and $S$ are chosen appropriately. 
This means that the first iteration of the ensemble MCMC, starting from an uninformative initial guess, can be run over a large number of steps ($l_1$) to ensure convergence within a relatively small amount of time.
On the other hand, each subsequent iteration starts from an ensemble of points which resembles the target distribution of that iteration, and hence requires fewer number of steps ($l_s$) to converge. 
Of course, the final iteration, over the true target distribution $P[\boldsymbol{\theta}|\mathfrak{D},\mathfrak{M}]$, should be run over enough steps ($l_S$) to ensure reliable inference and visualization of the distribution.
We stress here that only the final MCMC chain, obtained using the true target distribution $P[\boldsymbol{\theta}|\mathfrak{D},\mathfrak{M}]$, is used for inference, and the preceding iterations are performed only to aid convergence.

Comments are in order on why we randomly draw samples from the $s$th MCMC chain to use as starting points for the $(s+1)$th iteration (step 2b above), as opposed to starting the $(s+1)$th iteration from the end points of the $s$th MCMC chain.
Pulsar timing posterior distributions often contains regions which are difficult to sample efficiently, resulting in some of the MCMC chains getting stuck.
If we start the $(s+1)$th iteration from the end points of the $s$th MCMC chain, then any stuck chains from the $s$th iteration will definitely be passed on to the $(s+1)$th iteration.
On the other hand, by mixing the chains as done in step 2b, we reduce the probability of the stuck chain being passed on to a number less than 1.
Hence, as the number of iterations $S$ becomes large, the probability of a stuck chain surviving tends to 0, assuming that the number of stuck chains remains a sufficiently small fraction of the total number of chains in the MCMC ensemble.

Our scheme of tempering the posterior distribution by varying the data size is in contrast with the widely used likelihood tempering scheme \citep[e.g.][]{JohnsonMeyers+2024} where the intermediate  distributions are defined as 
\begin{equation}
P_\beta[\theta|\mathcal{D},\mathcal{L}]\propto P[\mathcal{D}|\theta,\mathcal{M}]^\beta P[\theta|\mathcal{M}]\,,
\label{eq:like-temper}
\end{equation}
where $\beta=[0,1]$ is an analogue of the inverse temperature in statistical physics.
There are two practical differences between the two schemes: (a) $\beta$ is a continuous parameter whereas the data size is by definition discrete; and (b) the cost of evaluating equation \ref{eq:like-temper} is independent of $\beta$ unlike in the data-tempering scheme.
The advantage of data size-tempering compared to $\beta$-tempering is the computational cost advantage, especially for large datasets.
The disadvantages is that the sub-datasets are randomly selected and can have posterior features which are absent in the full data-posterior.

An implementation of our method may be found in \url{https://github.com/abhisrkckl/vela-sdt}.

\section{Demonstration using PSR B1855+09}
\label{sec:demo}

We now proceed to demonstrate our method using the NANOGrav 12.5-year data of PSR B1855+09 \citep{AlamArzoumanian+2021}.
B1855+09 is a binary pulsar in a $\sim12$-day nearly circular orbit with a 0.25 solar mass white dwarf companion \citep{KaspiTaylorRyba1994}.

The dataset consists of 6464 narrowband TOAs obtained using the Arecibo telescope during a time span of $\sim12.5$ years (MJD 53358--57915).
The data were recorded using four different receiver-backend combinations, denoted as `430\_ASP', `430\_PUPPI', `L-wide\_ASP', and `L-wide\_PUPPI'.

The pulsar rotation is characterized by the rotational frequency and its derivative, together with a Fourier-basis Gaussian process with power law spectrum (with 16 harmonics) to characterize stochastic variations in the apparent spin frequency (achromatic red noise).
The solar system delays are parameterized by the ecliptic coordinates of the pulsar, proper motion, and parallax, and are computed using the DE436 solar system ephemeris.
Dispersion measure variations (dispersion noise) are modeled using a truncated Taylor series up to the second derivative, along with a Fourier-basis Gaussian process with a power law spectrum (with 32 harmonics).
The binary R{\o}mer and Shapiro delays are modeled using the ELL1 model accurate up to cubic terms in eccentricity \citep{LangeCamilo+2001,FioreLevin+2023}.
Instrumental delay between the `430' and `L-wide' receivers is represented as a phase jump.
Radiometer noise and pulse jitter noise are parameterized by the EFAC, EQUAD, and ECORR parameters for each receiver-backend combination \citep{ArzoumanianBrazier+2015}.
Finally, timing artifacts arising from frequency-dependent profile evolution are removed by fitting three `FD' parameters \citep{ArzoumanianBrazier+2015}.

The \texttt{Vela.jl} package is used to evaluate the log-posterior distribution(s).
The rotational frequency and its derivative, the sky coordinates, the dispersion measure and its derivatives, the phase jump, the `FD' parameters, and an overall phase offset are analytically marginalized assuming uninformative flat priors since these parameters behave either approximately or exactly linearly \citep{Susobhanan2025b}.
The Fourier amplitudes of the achromatic red noise and the dispersion noise are also analytically marginalized assuming Gaussian priors determined by the noise spectrum \citep{Lentati2013}.
The remaining timing and noise parameters together with their prior distributions are listed in Table 1.

Note that the analytic marginalization of the correlated noise amplitudes and the approximately linear parameters is necessary to avoid the Neal's funnel-like geometries which make the posterior intractable using most MCMC methods, in addition to reducing the dimensionality of the parameter space.
An alternative to analytic marginalization for this purpose is the standardizing parameter transformation described in \citep{GundersenVan_Haasteren+2026}.

\begin{table*}[]
\begin{tabular}{lll}
\hline
\textbf{Parameter(s)} & \textbf{Description and Unit}                               & \textbf{Prior Distribution} \\\hline
PX                    & Parallax, $\varpi$ (mas)                                              & $\Pi(\varpi) = 3 R_{\max}^{-3} \varpi^{-4}$                     \\
PMELONG               & Proper motion in ecliptic longitude, $\mu_\lambda$ (mas/yr)                & \S                       \\
PMELAT                & Proper motion in ecliptic latitude,  $\mu_\beta$ (mas/yr)                 & \S                       \\
PB                    & Binary orbital period, $P_b$ (d)                                           & \S                       \\
A1                    & Projected semi-major axis of the pulsar orbit, $a_1$ (lightsecond) & \S                       \\
M2                    & Companion mass, $m_2$ (solar mass)                                 & LogUniform[$10^{-9}$, 100]   \\
SINI                  & Sine-inclination,   $\sin\iota$                                         & $\Pi(\sin\iota) = \sin\iota (1 - \sin^2\iota)^{-1/2}$                     \\
TASC                  & Epoch of ascending node passage, $T_{\ascnode}$ (MJD)                       & Uniform[$-\hat{P}_b/2, \hat{P_b}/2$]               \\
EPS1                  & First Laplace-Lagrange parameter,  $\epsilon_1$                          &      \S                       \\
EPS2                  & Second Laplace-Lagrange parameter,  $\epsilon_2$                         &     \S                        \\
TNDMAMP               & Spectral log-amplitude of the dispersion noise,   $\log_{10}A_{\text{DMN}}$           &       Uniform[--18,--9]                      \\
TNDMGAM               & Spectral index of the dispersion noise,   $\gamma_{\text{DMN}}$                   &         Uniform[0,7]                     \\
TNREDAMP              & Spectral log-amplitude of the achromatic red noise, $\log_{10}A_{\text{ARN}}$          &         Uniform[--18,--9]                      \\
TNREDGAM              & Spectral index of the achromatic red noise,   $\gamma_{\text{ARN}}$                &      Uniform[0,7]                       \\
EFAC                  & TOA uncertainty multiplicative factor                       & LogNormal[0, 0.25]             \\
EQUAD                 & TOA error added in quadrature (us)                          &  LogUniform[0.001, 100]                           \\
ECORR                 & TOA observation-wise correlation parameter (us)             &  LogUniform[0.001, 100]                   \\\hline       
\end{tabular}
\caption{The list of explicitly sampled model parameters and their prior distributions. `\S' denotes `cheat' prior distributions, which are uniform distributions centered around the maximum-likelihood estimates with a width equal to 100 times the maximum-likelihood uncertainty (see Appendix B of \citep{Susobhanan2025a}). TASC is measured from the center of the data span (MJD 55637). The PX prior assumes that the pulsar is uniformly distributed within a sphere of radius $R_{\max}$ centered at the solar system barycenter (a uniform distribution is unphysical for this parameter). The SINI prior is defined such that the cos-inclination is uniformly distributed between 0 and 1. There are four sets of EFACs, EQUADs, and ECORRs, corresponding to each receiver-backend combination, with identical priors as listed above.
The rotational frequency and its derivative (F0, F1), the sky coordinates (ELAT, ELONG), the dispersion measure and its derivatives (DM, DM1, DM2), the instrumental phase jump (JUMP1), the FD parameters (FD1, FD2, FD3), and the overall phase offset (PHOFF) are analytically marginalized assuming an irregular flat prior, and are not listed in this table.
}
\label{tab:params}
\end{table*}

We perform Bayesian parameter estimation using the method described in section \ref{sec:sdt}, using the \texttt{emcee} package to run each ensemble sampling iteration.
We use $\xi=0.8$, but ensure that for each sub-dataset, there are at least 32 TOAs per receiver-backend combination; this leads to $S=23$.
We also ensure that each approximate model has at least 4 harmonics for both the achromatic red noise and the dispersion noise.
For each iteration, we use a total of 130 MCMC chains (5 times the number of free parameters) in the sampler ensemble.
The chain length for the first iteration is $l_1=10000$, for the final iteration is $l_S=3000$, and for the intermediate iterations is $l_s=1000$.
We choose the initial samples as $$\boldsymbol{\theta}_{w0}=0.7\,\boldsymbol{\theta}_\texttt{ml} + 0.3 \, \boldsymbol{\theta}_\texttt{prior,w}\,,$$ where $\boldsymbol{\theta}_\texttt{ml}$ represent the maximum-likelihood values present in the pulsar ephemeris provided with the dataset and $\boldsymbol{\theta}_\texttt{prior,w}$ are samples drawn from the prior distribution. 

\begin{figure}[ht!]
    \centering
    \includegraphics[width=1\linewidth]{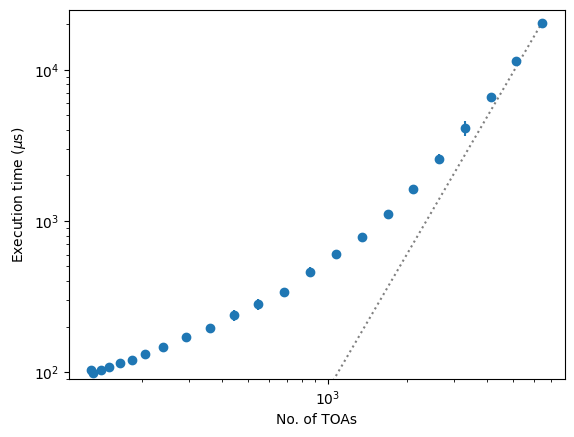}
    \caption{The execution time of the posterior distribution as a function of the number of TOAs ($n$). 
    The dotted line shows a function proportional to $n^3$.
    We see that the smaller data-subsets deviate significantly from the asymptotic $\mathcal{O}(n^3)$ behavior.}
    \label{fig:exec-time}
\end{figure}

The data-subset sizes and the corresponding log-posterior evaluation times are plotted in Figure \ref{fig:exec-time}.
We see that the plot is consistent with the expected $\mathcal{O}(n^3)$ asymptotic behavior for large data sizes, but deviate significantly for smaller datasets.
The full sequential MCMC run took 28 minutes on an AMD Ryzen 7 CPU, parallelized over 8 cores.
Out of this, the final iteration took approximately two-thirds of the total execution time; i.e., the preceding iterations aid the MCMC convergence without incurring too much computational cost. 

\begin{figure*}
    \centering
    \includegraphics[width=1\linewidth]{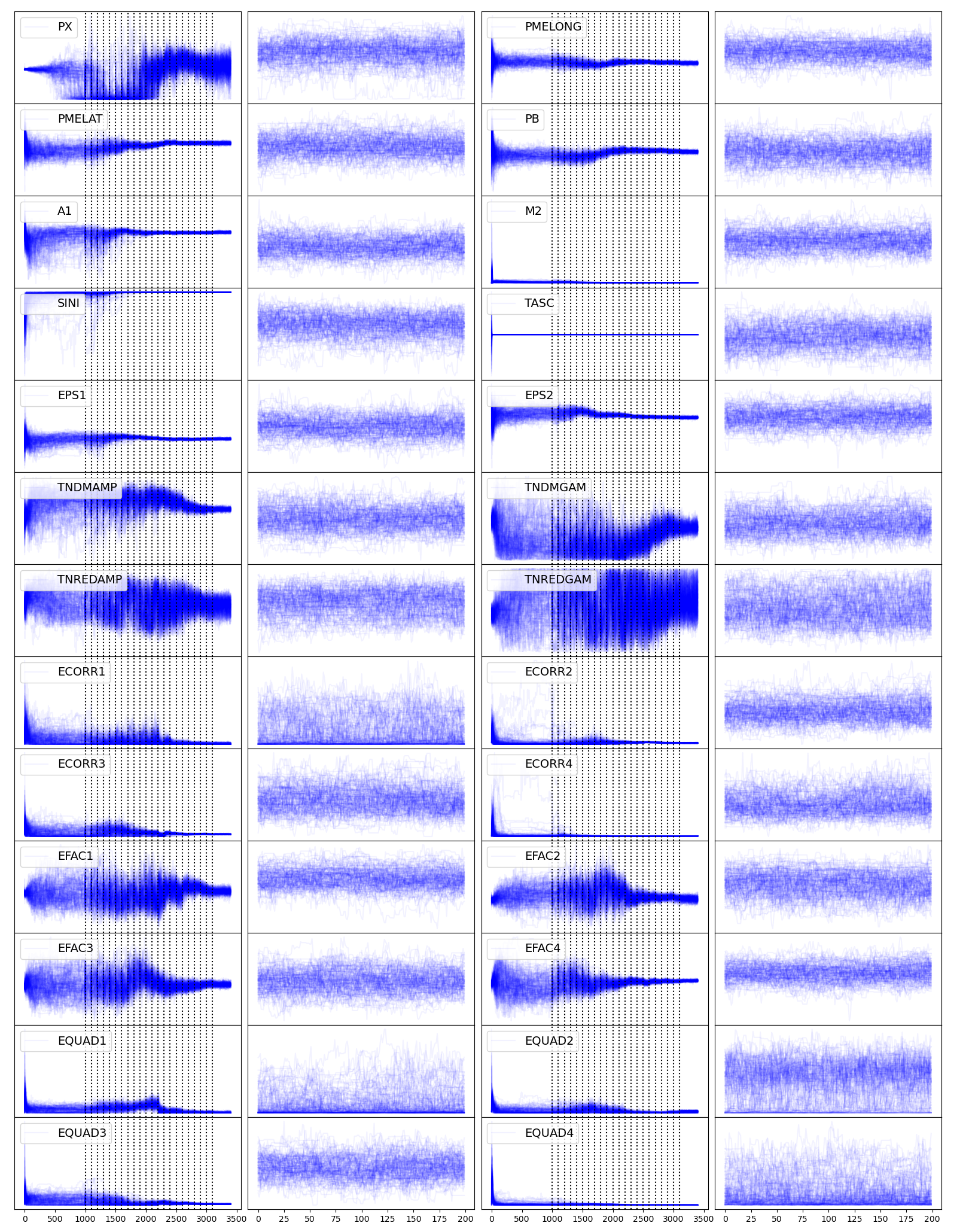}
    \caption{The concatenated MCMC chains for each parameter for all iterations, thinned by a factor of 10. The first and third columns show the full concatenated chain while the second and fourth columns show the corresponding burned-in final chain. The vertical dotted lines demarcate different iterations.
    The indices 1 to 4 of the EFACs, EQUADs, and ECORRs correspond to the reciever-backend combinations `430\_ASP', `430\_PUPPI', `L-wide\_ASP', and `L-wide\_PUPPI' in that order.
    See Table \ref{tab:params} for a description of each parameter.}
    \label{fig:sdt-chains}
\end{figure*}

The concatenated MCMC chains for each parameter from all iterations is plotted in Figure \ref{fig:sdt-chains}.
These plots illustrate how the parameters converge gradually as the iterations progress, as expected, indicating the robustness of the sampling scheme. 
Interestingly, these plots also indicate how the sensitivity of the experiment grows as more data is added, e.g., see the plot for the parallax (PX).

\begin{figure*}
    \centering
    \includegraphics[width=1\linewidth]{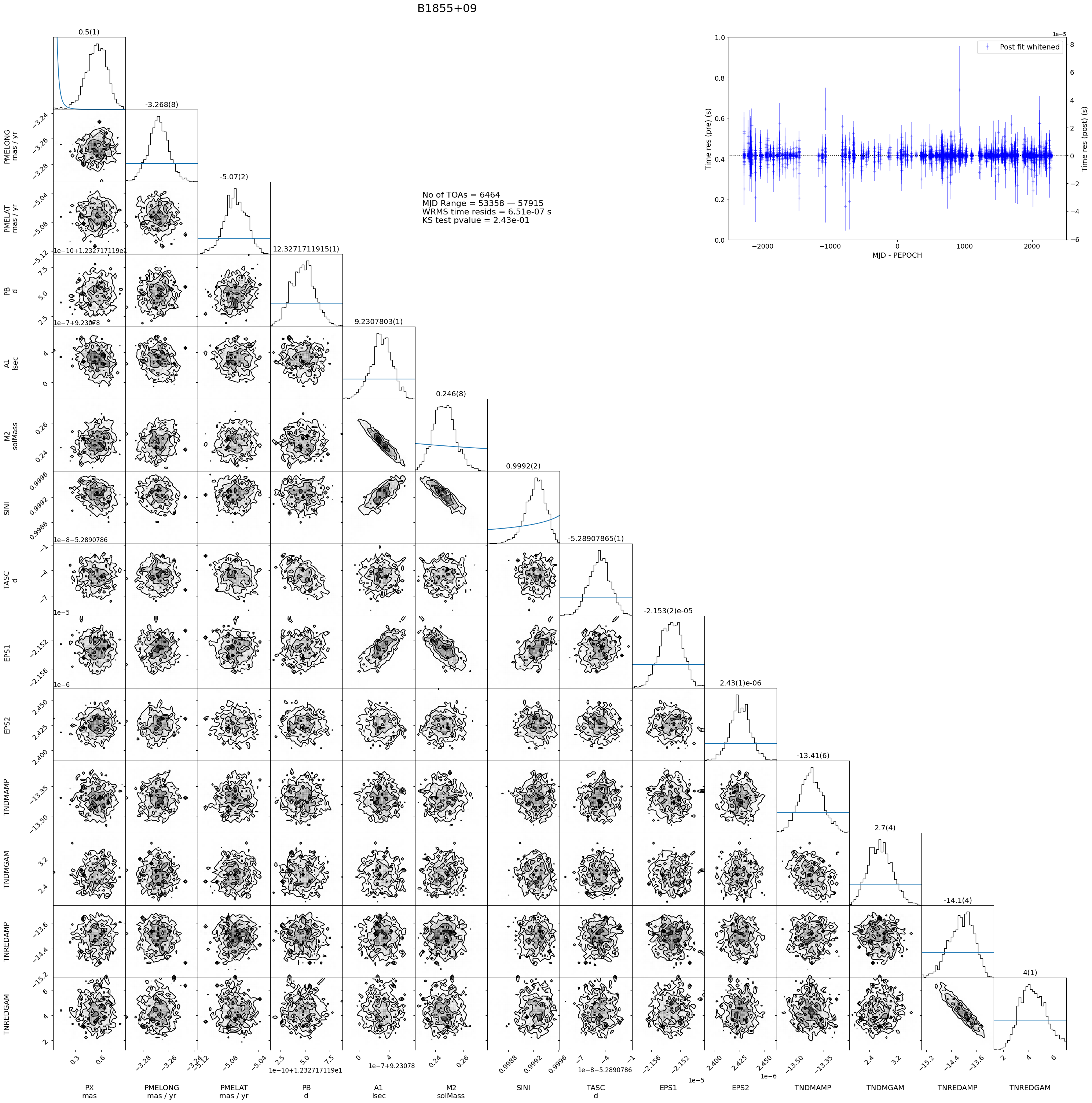}
    \caption{The corner plot of the target posterior distribution obtained from the final MCMC iteration.
    This plot corresponds to the final burned-in chain shown in the even columns of Figure \ref{fig:sdt-chains}.
    The prior distributions are plotted using blue lines. 
    The median parameter values along with the corresponding 1$\sigma$ uncertainties are shown above the 1-dimensional marginal distributions. 
    The inset shows the whitened post-fit residuals obtained using the median parameter values.
    The twelve white noise parameters (EFACs, EQUADs, and ECORRs) are not plotted due to space constraint.
    PEPOCH is the epoch at which the pulsar frequency and its derivatives are defined, chosen to be at the center of the data span (MJD 55637).
    TASC is measured from the PEPOCH.}
    \label{fig:post-corner}
\end{figure*}

Finally, a corner plot of the final posterior samples and the whitened post-fit residuals are shown in Figure \ref{fig:post-corner}.
Figure \ref{fig:post-corner} also lists the estimated value and uncertainty for each parameter.
We see that all parameters are well-constrained, although some parameters have significant covariances (e.g., M2 and SINI).
The whitened post-fit residuals do not show any un-modeled artifact, with a weighted root-mean-squared value of 0.6 $\mu$s. 
A Kolmogorov-Smirnov test of the whitened residuals for unit-Gaussianity reveals a $p$-value of 0.24, and suggests that our timing and noise model is  adequate for this dataset.

\section{Summary and Discussion}
\label{sec:summary}
We developed a sequential ensemble MCMC scheme for Bayesian non-linear pulsar timing and noise analysis where the data size is used as the tempering parameter.
Our method provides gradual and reliable convergence of the MCMC chains by iteratively increasing the data size and hence its sensitivity, starting from a possibly un-informed set of initial points, and gradually narrowing the target distribution until the  target posterior distribution of the full dataset is reached.
The initial iterations of our method are run using sub-datasets which can be orders of magnitude smaller than the full dataset, ensuring low computational cost.
The computational cost is further reduced by using approzimate lower-rank models derived from the target model, which is justified by the lower sensitivity of the sub-datasets.
The Bayesian inference is done using the MCMC chain from the final iteration corresponding to the true target distribution after applying appropriate burn-in.

We demonstrated the efficacy of our method using the NANOGrav 12.5-year data of PSR B1855+09, and showed that the MCMC chains gradually converge to the true target distribution as the iterations progress.
Further, we also demonstrated that the majority of the computational time is spent drawing samples from the true target distribution, rather than in the preceding iterations.

While we motivated and demonstrated this method in this paper using the case of Bayesian single-pulsar timing and noise analysis, it can be easily adapted to linearized Bayesian single-pulsar noise analysis following the exact same recipe.
On the other hand, a different strategy for creating data-subsets is required for applying our method to PTA gravitational wave background analysis, since the computational cost there is dominated by the evaluation of the Hellings-Downs correlations \citep{HellingsDowns1983} and the matrix operations involving them.
\citep{JohnsonMeyers+2024} showed that the computational cost thereof varies approximately quadratically with the number of pulsars.
Hence, an appropriate strategy for applying our method there would be to pick subsets of pulsars rather than subsets of TOAs.

We may also envisage parallel-tempering MCMC as well as normalizing flow-based sequential MCMC or variational inference methods where the data size is used as the tempering/annealing parameter rather than the inverse-temperature $\beta$. 
Such methods will of course have the advantage that the smaller dataset-posteriors are much cheaper to evaluate, leading to overall lesser computational cost.

\section*{Data availability}
The NANOGrav 12.5-year dataset is available at \url{https://nanograv.org/science/data/125-year-pulsar-timing-array-data-release}.

\section*{Software}
\texttt{Vela.jl} \citep{Susobhanan2025a,Susobhanan2025b}, \texttt{PINT} \citep{LuoRansom+2021,SusobhananKaplan+2024},
\texttt{emcee} \citep{Foreman-MackeyHogg2013},
\texttt{corner} \citep{Foreman-Mackey2016},
\texttt{numpy} \citep{HarrisMillman+2020},
\texttt{matplotlib} \citep{Hunter2007}.



\begin{acknowledgments}
We thank Rutger van Haasteren and Churchil Dwivedi for their comments on this manuscript.

During this work, we used ChatGPT for code review and minor grammatical and stylistic editing. All scientific analysis, interpretation, and conclusions were independently verified by the authors.
\end{acknowledgments}


\bibliography{vela-sdt}

\end{document}